\documentclass[aps,twocolumn,prd,superscriptaddress,showpacs,letterpaper,nofootinbib]{revtex4}
\usepackage{amsmath,amssymb}
\usepackage{color}
\usepackage[pdftex]{graphicx}
\usepackage[colorlinks=true,linkcolor=blue]{hyperref}
\usepackage{bm}

\makeatletter \newcommand{\rmnum}[1]{\romannumeral #1}
\newcommand{\Rmnum}[1]{\expandafter\@slowromancap\romannumeral #1@}

\begin{document} %%%%%% Title %%%%%%

\title{Sensitivity of intracavity filtering schemes for detecting
gravitational waves}

%%%%%% Authors %%%%%%

\author{Mengyao Wang} \affiliation{School of Physics and Astronomy,
University of Birmingham, Birmingham, B15 2TT, United Kingdom}

\author{Haixing Miao} \affiliation{Theoretical Astrophysics 350-17,
California Institute of Technology, Pasadena, California 91125, USA}

\author{Andreas Freise} \affiliation{School of Physics and Astronomy,
University of Birmingham, Birmingham, B15 2TT, United Kingdom}

\author{Yanbei Chen} \affiliation{Theoretical Astrophysics 350-17,
California Institute of Technology, Pasadena, California 91125, USA}

%%% Begin of Abstract %%%%%

\begin{abstract}
We consider enhancing the sensitivity of future
gravitational-wave detectors by adding optical filters inside the
signal-recycling cavity---an intracavity filtering scheme, which coherently
feeds the sideband signal back to the interferometer with a proper
frequency-dependent phase. We study three cases of such a scheme with
different motivations: (\rmnum{1}) the case of backaction noise
evasion, trying to cancel radiation-pressure noise with only
one filter cavity for a signal-recycled interferometer; (ii) the
speed-meter case, similar to the speed-meter scheme
proposed by Purdue and Chen [Phys. Rev. D {\bf 66}, 122004 (2002)] but without the 
resonant-sideband-extraction mirror, and also relieves the optical requirement on the 
sloshing mirror; (iii) the broadband detection
case with squeezed-light input,  numerically optimized for a
broadband sensitivity.
\end{abstract}

%%% End of Abstract %%%%%


\maketitle

%%%%%% Main body %%%%%%

\section{Introduction} With advanced gravitational-wave (GW)
detectors, such as Advanced LIGO\,\cite{Harry10} and Advanced
VIRGO~\cite{Accadia11}, now under construction, we will soon enter the
stage in which the quantum noise, arising from vacuum fluctuation of
the electromagnetic field, starts to play a significant role and sets
the sensitivity limit over most of the detection band. The GW
community has started a significant effort to develop
techniques for quantum-noise reduction\,\cite{Chen13,
Danilishin12}. Such techniques generally include (\rmnum{1})
modifying the input optics (input filtering), the use of
frequency-dependent squeezing realized by using filter cavities to
rotate the squeezing angle of the squeezed light in a
frequency-dependent way, which enables a simultaneous reduction of the
low-frequency radiation-pressure noise and the high-frequency shot
noise; (\rmnum{2}) modifying the output optics (output filtering),
the frequency-dependent readout by rotating the readout quadrature
angle with filter cavities, which allows us to measure the proper
quadrature and in turn cancel the radiation-pressure noise;
(\rmnum{3}) coherently feeding the GW signals back to the
interferometer, e.g., the signal-recycled Michelson\,\cite{Buonanno01}
creating an optical spring that
modifies the test mass dynamics, and the speed meter with a sloshing
cavity\,\cite{Purdue02} storing the earlier differential
displacement information of the test masses to ``slosh'' it back to the
interferometer with a negative sign or using an orthogonal polarized field
to cancel the position information\,\cite{Wade12}. 

\begin{figure}[!b]
\includegraphics[width=0.33\textwidth]{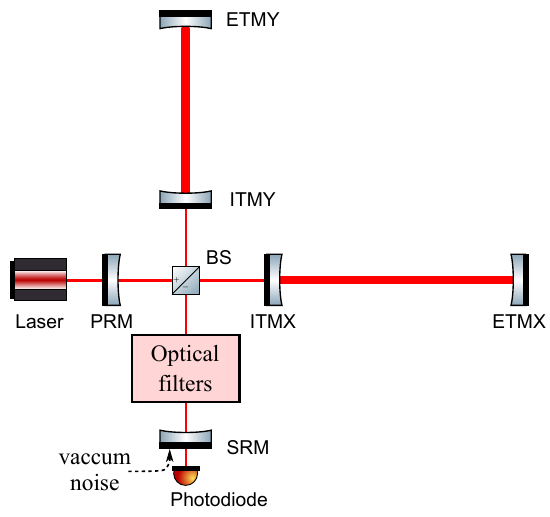}
\caption{The intracavity filtering scheme.
Here additional optical filters are introduced between the main interferometer
and the SRM. The vacuum noise enters from the
dark port, and the differential motion of the ITM
and end test mass (ETM) encodes the information of GWs. The power
recycling mirror (PRM) is used to coherently amplify the optical
power.
\label{fig:configuration}}
\end{figure}

Here we consider a new scheme in which we place the optical filters
inside the signal-recycling cavity, formed by the cavity input test
masses (ITMs) and the signal-recycling mirror (SRM), and which we
referred as an intracavity filtering
scheme. Figure \ref{fig:configuration} schematically illustrates the
configuration of a generic intracavity filtering scheme. The sideband fields are 
coherently fed back to the main interferometer. In some sense, this is one 
example of coherent feedback that has been extensively discussed in the 
quantum optics and control 
communities\,\cite{Gough2007,Mabuchi2008,Kerckhoff2010,Hamerly2012}. 
Depending on the choices of optical filters, different interferometer responses 
and sensitivities can be
obtained. We will not exhaust all the possibilities, and only focus on
the following three cases that have clear motivations.

Our first consideration is motivated by the frequency-dependent
readout scheme considered by Kimble \textit{et
al}~\cite{Kimble02}. Ideally, two filter cavities are able to
completely cancel the radiation-pressure noise for Advanced LIGO.
We investigate whether an intracavity filtering scheme with only one filter cavity
can provide the same radiation-pressure
noise cancellation or not. The answer turns out to be yes, but at a price of poorer 
sensitivity compared with frequency-dependent readout as shown in Sec.\,\ref{sec:caseI}.

Motivated by the speed-meter scheme with a
sloshing cavity proposed by Purdue and Chen\,\cite{Purdue02},
we investigate a scheme of using the intracavity filtering to create a speed
meter. The scheme is
similar to the sloshing speed meter but without the
resonant-sideband-extraction (RSE) mirror. Interestingly, we find that
the characteristic frequency $\omega_s$ for the speed response is given by
the geometric mean of the arm cavity bandwidth and the sloshing-cavity
bandwidth. This, as we will show in Sec.\,\ref{sec:caseII},
relieves the stringent requirement on the sloshing-cavity bandwidth in
the original design proposed in~\cite{Purdue02}.

Our third case is motivated by the goal of achieving a broadband
enhancement in the sensitivity based on current designs of the
advanced GW detectors, such as Advanced LIGO.
We numerically optimize the sensitivity of the intracavity
filtering scheme. With a reasonable specification for the optical loss and 
improvement of the classical noise, we obtain enhancement comparable to
frequency-dependent squeezing, and better than the
frequency-dependent readout. For the optimization, we use the cost
function introduced in~\cite{Miao13} which tries to maximize the
improvement over a broad band. The details are presented in
Sec.\,\ref{sec:caseIII}.

\section{Case I: Canceling radiation pressure noise}
\label{sec:caseI} In this section, we  first give a
brief review of the frequency-dependent readout for evading the
radiation-pressure noise presented in~\cite{Kimble02}, which can be
as a reference for analyzing alternative schemes, and present our
first case, using the
intracavity filtering scheme for canceling the radiation-pressure noise.

\subsection{Brief review of frequency-dependent readout scheme}
\label{subsec:frequency_dependent_readout}
The input-output relation for the amplitude and phase quadratures of a
general, {\it tuned} interferometer can be written as
\begin{equation} {\bm b}={\bf M} \, {\bm a} + {\bm D}\, h\,,
\label{eq:input_output_relation_compact}
\end{equation} and more explicitly, by expanding out the vectors $\bm
a$, $\bm b$, $\bm D$ and matrix $\bf M$, it is
 \begin{equation}\label{eq:input_output_relation}
\left[\begin{array}{c} b_1 \\ b_2
       \end{array} \right]= e^{2i\phi }\left[\begin{array}{cc} 1 & 0
\\ -{\cal K} & 1
       \end{array} \right] \left[\begin{array}{c} a_1 \\ a_2
               \end{array} \right]+e^{i\phi } \left[
 \begin{array}{c} 0 \\ \sqrt{2{\cal K}}
 \end{array} \right]\frac{h }{h_{\rm SQL} }\,.
 \end{equation}
Here $a_1\,(b_1)$ and $a_2\,(b_2)$ are the input (output) amplitude quadrature and phase quadrature, respectively,
which are functions of the sideband frequency $\Omega$ with respect to
the input laser frequency $\omega_0$; $\phi $ is the extra phase
factor; $h$ is the GW strain and $h_{\rm SQL}$ the standard quantum
limit (SQL) for the strain sensitivity given
by\,\cite{Braginsky92}
\begin{equation}\label{eq:SQL}
h_{\rm SQL}=\sqrt{\frac{8\hbar}{m\Omega^2 L_{\rm arm}^2}}
\end{equation}
with $L_{\rm arm}$ being the arm length and $m$ being
the mass of test masses (TMs); ${\cal K}$ quantifies the measurement strength which
is proportional to the optical power impinged onto the TMs and
also the mechanical response of the TMs and it is given by
\begin{equation}\label{eq:Kappa}
{\cal K}=\frac{16\omega_0\gamma I_c}{m L_{\rm arm} c\Omega^2(\Omega^2+\gamma^2)}
\end{equation}
for a tuned, signal-recycled Michelson interferometer,
with $I_c$ the optical power inside the arm cavities and $\gamma$
being the detection bandwidth jointly determined by the arm cavities
and signal-recycling cavity.

The homodyne readout allows us to measure a linear combination of the
output amplitude ($b_1$) and phase quadrature ($b_2$). When only the
phase quadrature is measured, the readout is given by (normalized with
respect to $h$)
\begin{equation}\label{eq:readout_phase_quadrature}
y =\frac{e^{i\phi} h_{\rm SQL}}{\sqrt{2}} \left(-\sqrt{\cal K}\, a_1 +
\frac{1}{\sqrt{\cal K}}a_2\right) + h\equiv \delta h + h\,.
\end{equation}
Here $\delta h$ is the strain-referred quantum noise
with the first term being the radiation-pressure noise (also termed as backaction) and the second
term being the shot noise. Its noise spectral density,\footnote{We
use single-sided spectral density defined as $\langle
\psi |A(\Omega) B(\Omega')+B(\Omega') A(\Omega)|
\psi\rangle=\frac{1}{2}S_{AB}\delta(\Omega-\Omega')$. Here
$|\psi\rangle$ is the quantum state for the optical field and the
vacuum state $|0\rangle$ is used for evaluating the quantum noise
which gives the cross spectral density $S_{11}=S_{22}=1$ and
$S_{12}=0$ among the input amplitude quadrature $a_1$ and phase
quadrature $a_2$.} for uncorrelated amplitude and phase vacuum noise,
can be obtained as
\begin{equation}\label{eq:spectral_density_phase_quadrature}
S_h =\left[\frac{\cal K}{2}+\frac{1}{2{\cal K}}\right] h_{\rm SQL}^2 \ge
h^2_{\rm SQL}\,.
\end{equation}
and is bounded by the SQL.

However, if a quadrature different from the phase quadrature is
measured, the SQL is no longer the limit. Furthermore, as shown
in~\cite{Kimble02}, measuring the quadrature with an angle
satisfying the following frequency dependence:
\begin{equation}\label{eq:back_action_evasion_condition}
\zeta =\arctan{\cal K}\,
\end{equation}
results in [see Eq.\,\eqref{eq:input_output_relation}]
\begin{equation}\label{eq:back_action_evasion_quadrature}
b_{\zeta}=b_1 \sin\zeta +b_2 \cos\zeta = \left( e^{2i\phi} a_2 +
e^{i\phi} \frac{\sqrt{2{\cal K}}h}{h_{\rm SQL}}\right)\cos\zeta\,.
\end{equation}
The radiation-pressure noise is therefore canceled,
leading to a sensitivity limited only by shot noise:
\begin{equation}\label{eq:back_action_evasion_spectral}
S_h^{\rm BAE}= \frac{h_{\rm SQL}^2}{2\cal K}\,.
\end{equation}

To achieve above frequency dependence, one can employ  a cascade of
filter cavities which rotates the quadrature depending on the frequency, as shown
in~\cite{Kimble02}. Specifically, the effects of filter cavities on
the quadratures can be described by the
following matrix:
\begin{equation}\label{eq:filter_cavity_rotation_matrix}
 {\bf M}_f =
e^{i \phi_f}\left[\begin{array}{cc} \cos\zeta_f & -\sin\zeta_f \\
\sin\zeta_f & \cos\zeta_f\end{array} \right]\,,
\end{equation}
 with the rotation angle $\zeta_f$ and the phase shift $\phi_f$ being
\begin{equation}\label{eq:zeta_alpha_relation}
\zeta_f=\frac{\alpha_++\alpha_-}{2}\,, \quad
\phi_f=\frac{\alpha_+-\alpha_-}{2}\,,
\end{equation}
where $\alpha_+$ and $\alpha_-$ are phase shifts of
upper and lower sidebands induced by the filter cavity.  In
particular, in the narrow band approximation ($\Omega \ll \omega_{\rm
fsr}$ with $\omega_{\rm fsr}$ being the free spectral range), for one
filter cavity we have
\begin{equation}\label{eq:upper_sideband_phase_shift_filter_cavity}
e^{i\alpha_\pm(\Omega)} =-\frac{\Omega\mp\Delta_f\mp i\gamma_f}{\Omega\mp\Delta_f\pm i\gamma_f}\,,
\end{equation}
with $\Delta_f$ and $\gamma_f$ being the cavity detuning and bandwidth, respectively.

A proper choice of the filter cavity parameters enables the desired
frequency-dependent rotation of the quadratures.  As proven in
Appendix A of~\cite{Purdue02} [see Eq.\,(A12) in their paper], when
the required $\tan\zeta$ of the rotation angle is a rational function
in $\Omega^2$ with the highest order of $\Omega^{2n}$, $n$ filter
cavities are needed. For obtaining $\tan\zeta={\cal K}$ in
Eq.\,\eqref{eq:Kappa}, which is a rational function of $\Omega^2$ with
the highest order being $\Omega^4$, two filter cavities are therefore
required.

\subsection{Intracavity filtering for canceling radiation pressure noise}
\label{subsec:intra_filtering_frequency_dependent_readout}

\begin{figure}[!b]
\includegraphics[width=0.48\textwidth]{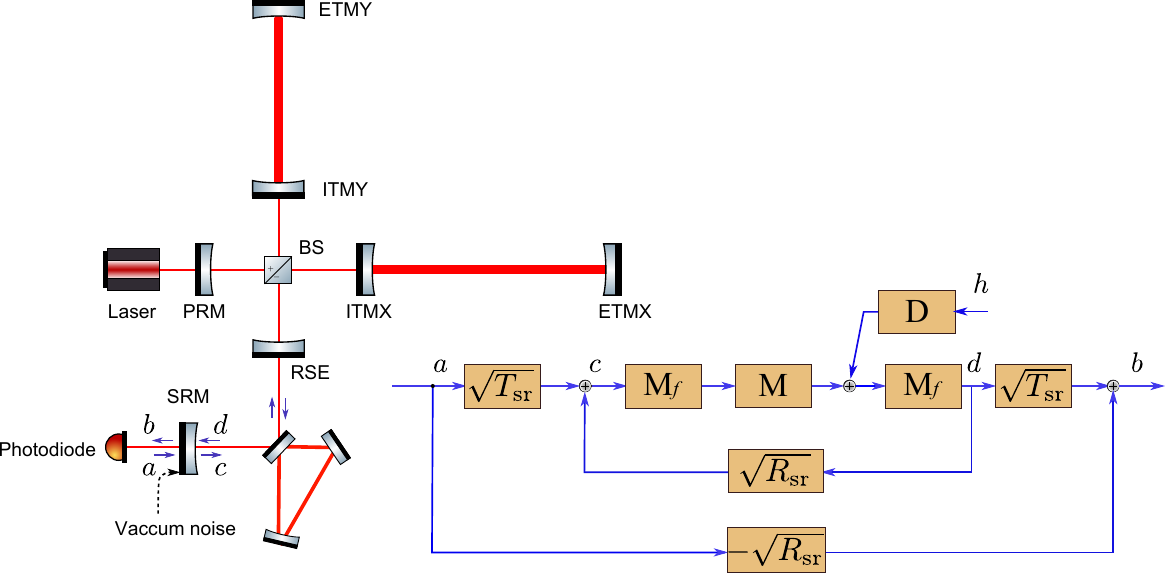}
\caption{An intracavity filtering scheme with a single filter together with a
RSE mirror for canceling the radiation-pressure noise (left) and its block 
diagram (right). Blocks ${\bf M,
M}_{f} $ and ${\bf D}$ are defined in
Eqs.\,\eqref{eq:input_output_relation_compact} and
\eqref{eq:filter_cavity_rotation_matrix}. $T_{\rm sr}$ and $R_{\rm
sr}$ are the transmissivity and reflectivity of the SRM.}
\label{fig:back_action_evasionI}
\end{figure}

In Fig.\,\ref{fig:back_action_evasionI} we show the corresponding
intracavity filtering scheme in which one filter cavity and a
resonant-sideband-extraction (RSE) mirror are placed inside the signal-recycling cavity. The RSE mirror
is used to effectively remove the frequency response of the arm cavities such that only 
one filter cavity is needed for canceling the radiation-pressure noise (the highest frequency 
dependence of ${\cal K}$ becomes $\Omega^2$ instead of $\Omega^4$). More explicitly, the
filter cavity sees an input-output relation for the optical field
in a form similar to Eq.\,\eqref{eq:input_output_relation}, but with
${\cal K}$ being replaced by
\begin{equation}\label{eq:little_kappa}
\kappa = \frac{8I_c
\omega_0}{m c^2 \Omega^2}\equiv \frac{\Omega_q^2}{\Omega^2}\,,
\end{equation}
where $\Omega_q$ is a characteristic frequency at which
the sensitivity curve touches the SQL.

To obtain the condition for canceling radiation-pressure noise, we first look at
the combined effect of $\bf M$ [see
Eq.\,\eqref{eq:input_output_relation}, replacing $\cal K$ by
$\kappa$] and ${\bf M}_f$ inside the signal-recycling cavity. It is described by ${\bf M}_{\rm tot}\equiv{\bf M}_f{\bf M}\,{\bf M}_f$ which reads
\begin{equation}\label{eq:Mf_M_effect}
{\bf M}_{\rm tot}=e^{2i\phi_{\rm tot}}
\left[\begin{array}{cc} \cos2\zeta_f + \frac{\kappa}{2}\sin2\zeta & -\sin2\zeta -\kappa \sin^2\zeta \\ \sin2\zeta_f -\kappa \cos^2\zeta_f& \cos2\zeta_f + \frac{\kappa}{2}\sin2\zeta_f \end{array} \right]
\end{equation}
with $\phi_{\rm tot}=\Omega L_{\rm arm}/c + \phi_f$. In order to remove the radiation-pressure noise 
from the phase quadrature as the usual frequency-dependent readout, we require the above matrix to be upper triangular, i.e.,
\begin{equation}
-\kappa \cos^2\zeta_f+\sin2\zeta_f =0
\end{equation} or equivalently,
\begin{equation}\label{eq:back_action_evasion_condition_intraI}
\tan\zeta_f = {\kappa}/{2}\,,
\end{equation}
which is achieved by choosing the filter cavity detuning and bandwidth to be
\begin{equation}\label{eq:filter_cavity_parameter}
\Delta_f =\gamma_f =
\,{\Omega_q}/{2}\,.
\end{equation}
Under this condition, the above matrix simply becomes
\begin{equation}\label{eq:Mtot_simplified}
{\bf M}_{\rm tot}=e^{2i\phi_{\rm tot}}\left[\begin{array}{cc} 1 & -\kappa \\ 0 & 1 \end{array} \right]\,.
\end{equation}
Since there is no additional rotation to ${\bf M}_{\rm tot}$ during the
propagation (see the block diagram in
Fig.\,\ref{fig:back_action_evasionI}), if we measure the output phase
quadrature $b_2$, which only depends on the input phase quadrature
$a_2$, we shall obtain a sensitivity only limited by the shot
noise. The final input-output relation is given by
\begin{equation}\label{eq:input_output_relation_intraI}
\bm b=\left[-\sqrt{R_{\rm sr}}\, {\bf I} +T_{\rm sr} {\bf M}_c{\bf M}_{\rm tot}\right]\bm a + \sqrt{T_{\rm sr}}\,{\bf M}_c{\bf M}_f {\bf D}
\,h\,,
\end{equation}
with ${\bf M}_c\equiv \left[{\bf I}-\sqrt{R_{\rm sr}}\,
{\bf M}_{\rm tot}\right]^{-1}$ and $\bf I$ the identity
matrix.

\begin{figure}[!t]
\includegraphics[width=0.43\textwidth]{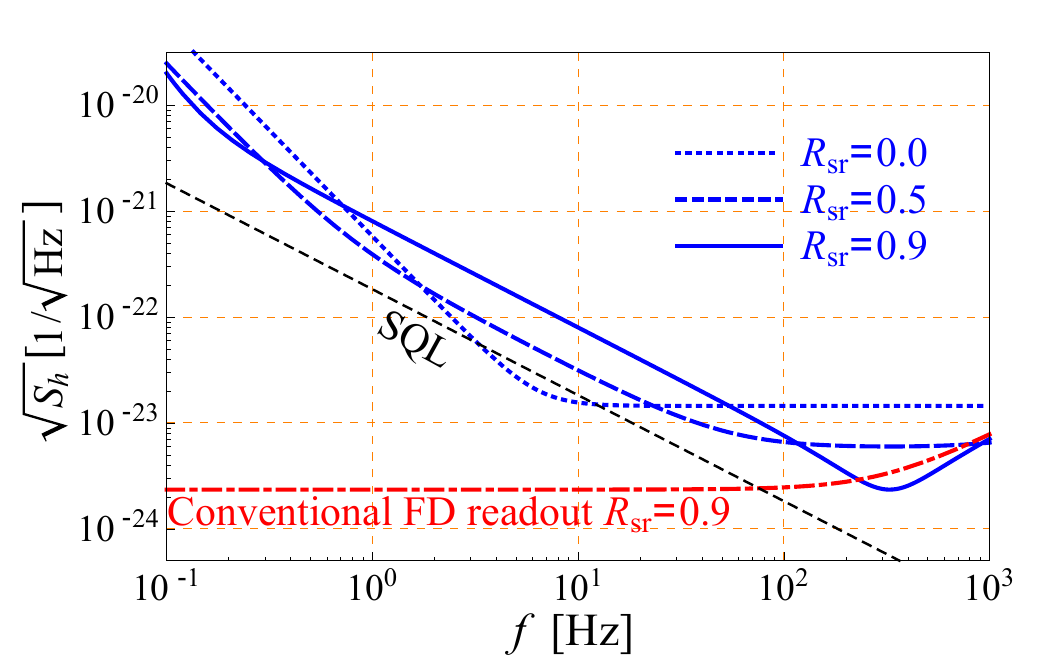}
\caption{The sensitivity of the intracavity filtering scheme for
evading radiation-pressure noise with different signal-recycling
mirror reflectivities (blue) in comparison with the conventional
frequency-dependent readout (FD) scheme proposed in~\cite{Kimble02} which uses two
additional filter cavities to filter the output from the signal-recycling mirror.
\label{fig:back_action_evasion_intra_sensitivity}}
\end{figure}

Naively one might expect that by just placing {\it one} filter cavity inside the
signal-recycling cavity we can achieve the same sensitivity as the frequency-dependent readout. However, as shown in
Fig.\,\ref{fig:back_action_evasion_intra_sensitivity}, this is not the
case, and the performance is poorer. Moreover, the sensitivity
at intermediate frequencies decreases as we increase the reflectivity of the signal-recycling mirror; 
even when the reflectivity goes to zero (no signal-recycling mirror), 
we do not recover the frequency-dependent readout.

To understand the sensitivity degradation
in comparison with the conventional frequency-dependent readout,
we first look at the case of $R_{\rm sr}=0$. We can write down
the input-output relation for the phase quadrature explicitly as (normalized with respect to strain)
\begin{equation}\label{eq:input_output_relation_phase_quadrature_intraI}
y_2 = \frac{e^{i\phi_{\rm tot}}h_{\rm SQL}}{\sqrt{2\kappa}\cos\zeta_f}a_2 +h\equiv \delta h + h\,.
\end{equation}
At low frequencies $\Omega\ll\Omega_q$, $\kappa\gg1$, $\cos\zeta_f\sim 1/\kappa$ and 
the strain-referred noise term $\delta h$ reads
\begin{equation}\label{eq:input_output_relation_phase_quadrature_intra_approximation}
\delta h|_{\Omega\ll\Omega_q}\approx \sqrt{\frac{\kappa}{2}}\,h_{\rm SQL} a_2 \propto \frac{h_{\rm SQL}}{\Omega}a_2\,.
\end{equation}
Therefore, even though it is a shot-noise limited sensitivity, the spectrum of the
shot noise is not flat and increases at low frequencies as shown by the dotted
line in Fig.\,\ref{fig:back_action_evasion_intra_sensitivity}. This comes from the
additional rotation of the input vacuum field by the filter cavity, which is absent in
the usual frequency-dependent readout [see
Eq.\,\eqref{eq:back_action_evasion_quadrature}].

For $R_{\rm sr}\neq 0$, the expression for the output phase quadrature (strain-referred) is:
\begin{equation}\label{eq:output_phase_quadrature}
y_2 =
\frac{e^{i\phi_{\rm tot}}(1-\sqrt{R_{\rm sr}}e^{2i\phi_{\rm tot}})h_{\rm
SQL}}{\sqrt{2T_{\rm sr}\kappa}\,\cos\zeta_f}\, a_2+h\equiv \delta h+h\,,
\end{equation}
with the phase factor being
\begin{equation}\label{eq:phi_total_expression}
e^{2i\phi_{\rm tot}} = e^{2i\Omega L_{\rm arm}/c} \frac{(\Omega-i\Omega_q/2)^2-\Omega_q^2/4}{(\Omega+i\Omega_q/2)^2-\Omega_q^2/4}\,.
\end{equation}
To understand the behavior as shown in
Fig.\,\ref{fig:back_action_evasion_intra_sensitivity},
we consider the case of $T_{\rm sr}\ll 1$ at three
different frequency regimes: (i) {\it at very low frequencies} $\Omega\ll \Omega_q/2$,
we have $e^{2i\phi_{\rm tot}}\sim 1$ and $\cos\zeta\sim 1/\kappa$. We can therefore obtain
\begin{equation}\label{eq:h_low_frequencies}
\delta h|_{\Omega \ll \Omega_q/2}\approx \sqrt{\frac{T_{\rm sr}\kappa}{8}} h_{\rm SQL}a_2\propto \sqrt{T_{\rm sr}}\frac{h_{\rm SQL}}{\Omega}a_2\,.
\end{equation}
The frequency dependence is the same as $R_{\rm sr}=0$ 
[see Eq.\,\eqref{eq:input_output_relation_phase_quadrature_intra_approximation}] 
but with an additional factor $\sqrt{T_{\rm sr}}$---the smaller $T_{\rm sr}$ the better 
the sensitivity; (ii) {\it at intermediate frequencies} around $\Omega_q/2$, we have
\begin{equation}\label{eq:phase_intermidiate_frequencies}
1-\sqrt{R_{\rm sr}}e^{2i\phi_{\rm tot}}\approx \frac{2i\,\Omega\,\Omega_q}{(\Omega+i\Omega_q/2)^2-\Omega_q^2/4}\,.
\end{equation}
At frequencies smaller than (yet still around) $\Omega_q/2$, it is approximated to be $-4i\Omega/\Omega_q$; 
at higher frequencies $\Omega\gtrsim \Omega_q/2$, it approximately equals to $2i\Omega_q/\Omega$. 
Therefore, we obtain the strain-referred noise term:
\begin{equation}\label{eq:h_intermediate_frequencies1}
\delta h|_{\Omega\lesssim \Omega_q/2}\propto \frac{\Omega \sqrt{\kappa}}{\sqrt{T_{\rm sr}}} \, h_{\rm SQL} a_2\propto \frac{h_{\rm SQL}}{\sqrt{T_{\rm sr}}}a_2\,,
\end{equation}
and
\begin{equation}\label{eq:h_intermediate_frequencies2}
\delta h|_{\Omega\gtrsim \Omega_q/2}\propto  \frac{1}{\Omega \sqrt{\kappa}\sqrt{T_{\rm sr}}} \, h_{\rm SQL} a_2\propto \frac{h_{\rm SQL}}{\sqrt{T_{\rm sr}}}a_2\,,
\end{equation}
where we used the fact that $\Omega\sqrt{\kappa}\propto\Omega^0$.  This explains why 
the spectrum is parallel to the SQL around intermediate frequencies. We also notice that 
the sensitivity decreases as we increase the reflectivity (smaller $T_{\rm sr}$); 
(iii) {\it at very high frequencies} $\Omega\gg \Omega_q/2$, we have
\begin{equation}\label{eq:phase_high_frequencies}
1-\sqrt{R_{\rm sr}}e^{2i\phi_{\rm tot}}\approx T_{\rm sr}/2-2 i\Omega L_{\rm arm}/c,
\end{equation}
and
\begin{equation}\label{eq:h_high_frequencies}
\delta h|_{\Omega \gg \Omega_q/2} \propto \frac{{T_{\rm sr}-4i\Omega L_{\rm arm}/c}}
{\sqrt{T_{\rm sr}\kappa}}h_{\rm SQL} a_2\,.
\end{equation}
At very high frequencies, the noise spectrum increases as frequency $\Omega$, 
which matches the spectrum behavior as shown in 
Fig.\,\ref{fig:back_action_evasion_intra_sensitivity} (the blue solid curve).

It is worthy mentioning that if the filter cavity is tuned to be resonant, instead of detuned for
evading the radiation-pressure noise, the above scheme becomes a speed
meter, even when the RSE mirror is removed, as we will discuss in Sec.\,\ref{sec:caseII}.

\section{Case II: Realizing a speed meter}
\label{sec:caseII}
In this section, we investigate the intracavity filtering scheme as a speed-meter, 
inspired by the speed meter scheme proposed by Purdue and Chen\,\cite{Purdue02}, 
which introduced an additional sloshing cavity to create speed response.

\subsection{Brief review of the speed meter with sloshing cavity}
\begin{figure}[!t]
\includegraphics[width=0.48\textwidth]{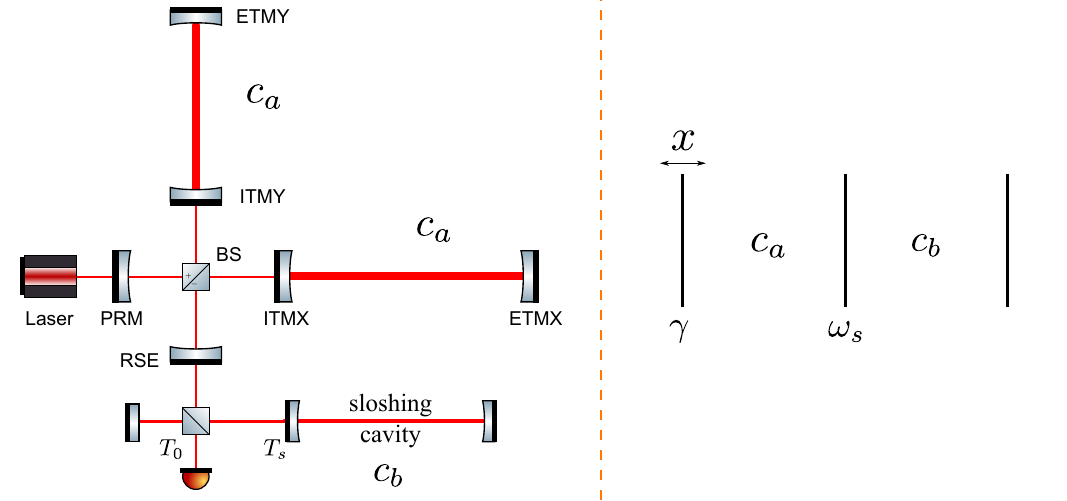}
\caption{The speed meter realized by adding an additional sloshing
cavity at the output port proposed in~\cite{Purdue02} (left) and its
simplified two-cavity-mode model (right).
\label{fig:speed_meter_sloshing_cavity}}
\end{figure}

The corresponding speed-meter scheme is
shown in Fig.\,\ref{fig:speed_meter_sloshing_cavity}, where a sloshing
cavity combined with a RSE mirror is added to the interferometer
output. Again the RSE mirror is applied to cancel the effect of the
ITMs of the arm cavities, and it has the same transmissivity as the
ITMs.
In this case the speed
response then can be understood qualitatively by using the model of
two coupled cavity modes as shown in the right part of
Fig.\,\ref{fig:speed_meter_sloshing_cavity}. In particular, the cavity
mode $c_a$ corresponds to the optical field inside the arm cavities,
and the cavity mode $c_b$ is the field inside the sloshing
cavity. These two are coupled via the sloshing mirror with a
characteristic coupling rate given by the sloshing frequency
$\omega_s$, which is defined as
\begin{equation}\label{eq:sloshing_frequency}
\omega_s=\frac{c\sqrt{T_s}}{2\sqrt{L_{\rm arm}L_s}}\,,
\end{equation}
with $L_s$ being the length of the sloshing cavity and
$T_s$ being the transmissivity of the sloshing mirror.

The classical equations of motion for these two cavity modes can be
written as
\begin{align}\label{eq:equation_of_motion_two_modes}
\dot c_a +\gamma c_a &=-i\, G\, x - i\,\omega_s c_b\,, \\ \dot c_b &= -i\,\omega_s
c_a\,.
\end{align}
Here $\gamma=4 c T_0/L_{\rm arm}$ is the signal extraction
rate and $G$ quantifies the response of the cavity mode to
the test mass displacement. Solving these two equations in the
frequency domain yields
\begin{equation}\label{eq:solution_to_ca}
c_a(\Omega) =
\frac{G\,\Omega }{\Omega^2-\omega_s^2 + i\Omega \gamma}x(\Omega)\,.
\end{equation}
At low frequencies $\Omega\ll \omega_s$ and a small
extraction rate $\gamma< \omega_s$, we obtain
\begin{equation}\label{eq:solution_to_ca_low_frequency} c_a(\Omega)
\approx -\frac{G\,\Omega}{\omega_s^2} x(\Omega)\propto -i\,\Omega
\,x(\Omega)\,,
\end{equation}
which implies a speed response.

The exact input-output relation for such a scheme is given
in~\cite{Purdue02} (see Eqs.\,(12), (13) and (14) in the paper), and
have a similar form as Eq.\,\eqref{eq:input_output_relation} with
$\cal K $ given by
\begin{equation}\label{eq:Kappa_beta_speed_meter}
{\cal K}_{\rm sm} =
\frac{16\omega_0 I_c}{m L_{\rm arm} c |\Omega^2-\omega_s^2+i\gamma
\Omega|^2}\,.
\end{equation}
Notice that ${\cal K}_{\rm sm}$ is nearly a constant at
low frequencies instead of having a strong frequency dependence. This
means that in order to evade the radiation-pressure noise at low
frequencies by satisfying
Eq.\,\eqref{eq:back_action_evasion_condition}, one can simply measure
a quadrature that is frequency independent
\begin{equation}\label{eq:back_action_evasion_condition_speed_meter}
\zeta_{\rm sm} = \arctan {\cal K}_{\rm sm}|_{\Omega\rightarrow 0}\,.
\end{equation}
A frequency-dependent readout is only needed when a better
sensitivity is required particularly at high frequencies. This is due
to the fact that high-frequency sensitivity is normally degraded if
$\zeta_{\rm sm}$ is different from zero (which represents
the phase quadrature).

\subsection{Intracavity filtering as a speed meter}

\begin{figure}[!b]
\includegraphics[width=0.48\textwidth]{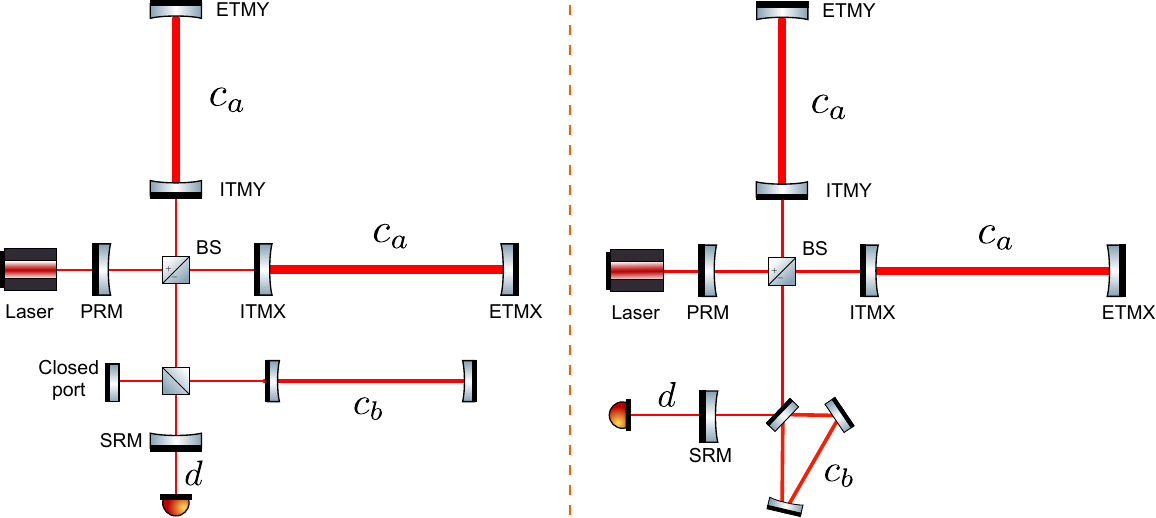}
\caption{Two equivalent intracavity filtering schemes as speed meter. Compared with the speed meter with a sloshing cavity shown in
Fig.\,\ref{fig:speed_meter_sloshing_cavity}, there is no RSE mirror
but a signal-recycling mirror (SRM).
\label{fig:speed_meter_intra_cavity}}
\end{figure}

Figure \ref{fig:speed_meter_intra_cavity} shows two equivalent
intracavity filtering speed-meter schemes without the RSE mirror.
In terms of complexity, it is the same as the previous
sloshing-cavity scheme, but the sensitivity performance is
different. As we will show, it has two interesting features. The {\it
first} one is that it also has a speed response and the characteristic
frequency $\omega_s$, up to which the speed response dominates, is
given by
\begin{equation}\label{eq:sloshing_frequency_intra_cavity}
\omega_s =\frac{c\sqrt{T_{\rm ITM} T_s}}{2\sqrt{L_{\rm
arm}L_s}}=\sqrt{\gamma_{\rm arm}\gamma_s}\,.
\end{equation}
This differs from Eq.\,\eqref{eq:sloshing_frequency} by
an extra factor of $\sqrt{T_{\rm ITM}}$, the transmissivity
coefficient of the arm cavity ITMs. Thus the sloshing frequency
is determined by the compound mirror formed by the ITMs and the
sloshing mirror. This feature makes it appealing in the sense that we
can realize a speed meter with a relatively short sloshing cavity. For
example, given the ITM transmittance $T_{\rm ITM}=0.01$ and
$T_s=900$ppm, we can set the sloshing frequency around 100Hz for a
100m sloshing cavity. However, this is very challenging for the speed
meter proposed in~\cite{Purdue02}, in which case the sloshing mirror
transmittance $T_s$ needs to be $90$ppm given a 100m sloshing cavity.
The {\it second} interesting feature is that it can also have a
position response at low frequencies when the parameters are chosen
properly. This can possibly provide a way to create a \textit{Local Oscillator}
(LO) for a practical readout scheme similar to the DC readout
realization described in~\cite{Wang13}.

To perform a detailed analysis of the scheme's quantum noise, one can use the
standard input-output formalism by writing down the propagation
equations for the fields and solve a set of linear equations in the
frequency domain. Instead, here we will follow the approach
given in~\cite{Buonanno03} by mapping parameters of the optics into several
characteristic quantities, and using the narrow band approximation to
define some effective modes. The advantage of this method is that it allows us
to gain a clearer insight into the dynamics of the intracavity
filtering scheme.  We define: (\rmnum{1}) $c_a$ --- the differential
mode of the two arm cavities; (\rmnum{2}) $c_b$ --- the cavity mode
inside the sloshing cavity; (\rmnum{3}) $d$ --- the external field;
and (\rmnum{4}) $\omega_s, \gamma_a$ and $\gamma_b$ --- the
characteristic frequencies for the coupling between $c_a, c_b$ and
$d$, as illustrated schematically in Fig.\,\ref{fig:speed_meter_intra_cavity}. These characteristic
frequencies, $\omega_s$, $\omega_a$, $\omega_b$, $\gamma_a$and
$\gamma_b$, are related to the parameters of the optical components,
which are shown explicitly in Appendix\,\ref{Appendix:A}.

We can then write down the Hamiltonian for the intracavity filtering
scheme consisting of two optical modes and one test mass with reduced
mass $m$, which reads
\begin{equation}\label{eq:Hamiltonian} \hat {\cal H}=\hat {\cal
H}_0+\hat {\cal H}_{\rm int}+\hat {\cal H}_{\rm ext}+\hat{\cal H}_{\rm
GW}\,.
\end{equation}
It contains four parts:

\noindent (i) The free Hamiltonian $\hat {\cal H}_0$ reads
\begin{equation}\label{eq:Hamiltonian_free_part}
\hat {\cal H}_0
=\hbar \omega_a\hat c_a^{\dag}\hat c_a +\hbar \omega_b \hat c_b^{\dag}\hat
c_b + \frac{\hat p^2}{2m}\,.
\end{equation}

\noindent(ii) The interaction Hamiltonian $\hat {\cal H}_{\rm int}$ is
\begin{equation}\label{eq:Hamiltonian_interaction_part}
\hat {\cal
H}_{\rm int}=\hbar G_a \hat x \hat c_a +\hbar \omega_{s}\hat
c_a^{\dag}\hat c_b + {\rm h.c.}\,,
\end{equation} where `h.c.' denotes the \textit{Hermitian
conjugate}. The first term represents the interaction between the cavity mode
$\hat{c}_a$ and the test mass $\hat x$ via radiation pressure with
$G_a=\omega_0\bar c_a/L_{\rm arm}$ and $\bar c_a$ being the
steady-state amplitude of $\hat c_a$ due to the coherent pumping of
the laser. The second term describes the coupling between the two cavity modes
with the coupling rate given by $\omega_{s}$.

\noindent (iii) The interaction Hamiltonian $\hat {\cal H}_{\rm ext}$
between the cavity modes and the external field $\hat d$ reads
\begin{equation}\label{eq:Hamiltonian_external} \hat {\cal H}_{\rm
ext}= i\hbar (\sqrt{2\gamma_a}\,\hat a^{\dag}+\sqrt{2\gamma_b}\,\hat
b^{\dag})\hat d \, e^{-i\omega_0 t} + {\rm h.c.}\,.
\end{equation}

\noindent(iv) The energy $\hat {\cal H}_{\rm GW}$ from the interaction
between the test mass and the GW tidal force $F_{\rm GW}$ is given by
\begin{equation}\label{eq:Hamiltonian_gravitational_wave} \hat {\cal
H}_{\rm GW}= -\hat x \, F_{\rm GW}\,.
\end{equation}

Given the above Hamiltonian, we can obtain the corresponding equations
of motion. Specifically, for the test mass, we obtain
\begin{equation}\label{eq:equation_of_motion_test_mass} m\ddot {\hat
x} =\hat F_{\rm rad} + F_{\rm GW}\,,
\end{equation} with the radiation-pressure force $\hat F_{\rm rad}$
defined as
\begin{equation}\label{eq:radiation_pressure_force} \hat F_{\rm
rad}\equiv -\hbar G_a(\hat c_a+\hat c_a^{\dag})\,.
\end{equation}
For the cavity modes $\hat c_a$ and $\hat c_b$, we have
\begin{align}\label{eq:equation_of_motion_cavity_modes} \dot {\hat
c}_a+(\gamma_a + i\Delta_a)\hat c_a & =-i G_a \hat x -i\omega_s \hat
c_b+\sqrt{2\gamma_a}\,\hat d_{\rm in}\,, \\ \dot{\hat
c}_b+(\gamma_b+i\Delta_b)\hat c_b & =-i\omega_s\hat c_a
+\sqrt{2\gamma_b}\,\hat d_{\rm in}\,.
\end{align}
The above detuning frequency $\Delta_{a,b}$ is defined as
$\Delta_j\equiv \omega_j-\omega_0\;(j=a, b)$.  The interferometer
output is related to the cavity modes through the standard
input-output relation:
\begin{equation}\label{eq:input_output_relation_Hamiltonian}
\hat d_{\rm out}=-\hat d_{\rm in}+\sqrt{2\gamma_a}\,\hat c_a +
\sqrt{2\gamma_b}\,\hat c_b\,.
\end{equation}

These equations can be solved in the frequency domain, and are
generally quite lengthy but straightforward. Here we focus on the
tuned case of $\Delta_a=\Delta_b=0$, which gives
\begin{align}
\hat c_a&=\frac{G_a(\Omega+i\gamma_b)\hat
x+[\sqrt{2\gamma_a}(\gamma_b-i\Omega)-i\sqrt{2\gamma_b}\omega_s]\hat
d_{\rm in}} {(\Omega+i\gamma_a)(\Omega+i\gamma_b)-\omega_s^2}\,,\\
\hat c_b&=\frac{G_a\omega_{s}\hat
x+[\sqrt{2\gamma_b}(\gamma_a-i\Omega)-i\sqrt{2\gamma_a}\omega_s]\hat
d_{\rm in}}{(\Omega+i\gamma_a)(\Omega+i\gamma_b)-\omega_s^2}\,.
\end{align}
and the radiation-pressure noise reads:
\begin{equation}\label{eq:F_rad_tuned_case_intra_cavity}
\hat F_{\rm
rad}=\frac{2\hbar G_a [\sqrt{\gamma_a}(\gamma_b-i\Omega)\hat
d_1+\sqrt{\gamma_b}\omega_{s} \hat
d_2]}{(\Omega+i\gamma_a)(\Omega+i\gamma_b)-\omega_s^2}\,,
\end{equation}
with $\hat d_1\equiv (\hat d_{\rm in}+\hat d_{\rm
in}^{\dag})/\sqrt{2}$ and $\hat d_2\equiv (\hat d_{\rm in}-\hat d_{\rm
in}^{\dag})/\sqrt{2}i$.

The response to the test-mass displacement at the output reads:
\begin{equation}\label{eq:output_displacement_response} \hat d_{\rm
out}^{|x} (\Omega)=
\frac{G_a[\sqrt{2\gamma_a}\,\Omega+\sqrt{2\gamma_b}(i\sqrt{\gamma_a\gamma_b}+\omega_s)]\hat
x(\Omega)}{(\Omega+i\gamma_a)(\Omega+i\gamma_b)-\omega_s^2}\,.
\end{equation}
As we can see, the first term in the bracket of the
numerator is proportional to $\Omega$, which gives the speed response,
while the remaining term proportional to $\sqrt{\gamma_b}$ gives the
linear displacement response. Therefore, it contains a mix of speed and
displacement response. This is similar to the polarizing Sagnac
interferometer with imperfect polarizing beam
splitter~\cite{Wang13}, and it implies a potential local oscillator for a homodyne
detection can be extracted. The turning frequency
$\Omega_{\rm turn}$, at which the speed response becomes dominant, is
given by [see Eq.\,\eqref{eq:output_displacement_response}]:
\begin{equation}
\Omega_{\rm turn}=\omega_s\sqrt{\frac{\gamma_b}{\gamma_a}}\,.
\end{equation}
In the limit of $\gamma_b\rightarrow 0$, it
approaches the speed meter
\begin{equation}\label{eq:output_displacement_response_gamma_b_0}
\hat d^{|x}_{\rm out}(\Omega)|_{\gamma_b\rightarrow 0}=\frac{\sqrt{
2\gamma_a}\,G_a \,\Omega }{\Omega^2+i\gamma_a\Omega-\omega_s^2}\hat
x(\Omega)\,,
\end{equation}
with a sloshing frequency given by
Eq.\,\eqref{eq:sloshing_frequency_intra_cavity} (with proper phase
chosen (see its definition in
Eq.\,\eqref{eq:sloshing_frequency_definition}).

\begin{figure}[!t]
\includegraphics[width=0.43\textwidth]{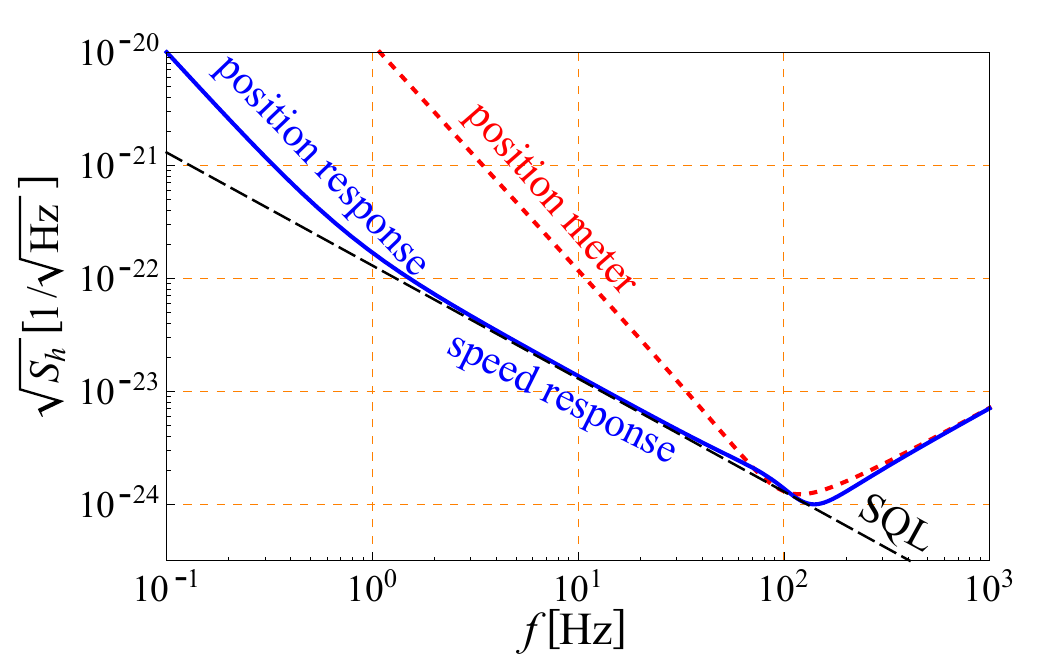}
\caption{The sensitivity curve for the intracavity filtering scheme
(blue solid) in comparison with the conventional position meter (red dotted). There
is a smooth transition from the position response to the speed response.
\label{fig:speed_meter_intra_cavity_sensitivity}}
\end{figure}

In Fig.\,\ref{fig:speed_meter_intra_cavity_sensitivity},
we show the resulting sensitivity by choosing
$\gamma_a/2\pi=\omega_s/2\pi=100\,\rm Hz$ and $\gamma_b/2\pi = 0.007\,\rm
Hz$ (blue solid). Indeed, at low frequencies the sensitivity
curve is similar to that of a position meter (red dotted) and it smoothly transits to the
speed-meter sensitivity in the intermediate frequencies. Given those parameters, the turning frequency $\Omega_{\rm turn}$ is around 1\,Hz which matches the blue curve.

These characteristic frequencies can be mapped to parameters for the optics by 
using their definitions in Eqs.\,\eqref{eq:sloshing_frequency_definition} and
\eqref{eq:decay_rate_definition} (Appendix\,\ref{Appendix:A}). We
summarize these parameters in Table\,\ref{tab:three_port}. Two possible designs are presented in 
terms of including
the SRM or not. From the mirror parameters shown in Table\,\ref{tab:three_port}, we conclude that 
an intracavity filtering scheme could be implemented as an alternative speed meter without stringent 
design requirements. Meanwhile, a scheme in the absence of the SRM could also have a speed-meter 
response, however, requiring a high reflection sloshing mirror.

\begin{table}[!b]
\begin{center}
\begin{tabular}{l | cc}
\hline \hline
Mirror$^{\dagger}$ & With SRM & Without SRM \\
\hline
 Sloshing mirror  & $0.0048\, (0.0)$ & $0.00080\, (0.0)$\\
 ITM  & $0.012\,(\pi)^{\ddagger}$ & $0.068\, (\pi)$ \\
 SRM & $0.50\,(0.0)$ &$1.0\,(0.0)$ \\
\hline
\end{tabular}
\end{center}
\caption{A table showing the power transmissivity (and the reflection phase) of relevant mirrors in 
the three-port junction as shown in
Fig.\,\ref{fig:three_port} (Appendix\,\ref{Appendix:A}), resulting in the sensitivity (blue) in
Fig.\,\ref{fig:speed_meter_intra_cavity_sensitivity}.
The left column values
refer to a model combining the SRM, the sloshing mirror, and the arm cavity ITMs.
The right-hand column values correspond to
a design without the SRM, however, giving the same sensitivity.\\
$^{\dagger}$Here we only specify the parameters for the left scheme in 
Fig.\,\ref{fig:speed_meter_intra_cavity}.\\
$^{\ddagger}$The phase of the ETMs needs to be $\pi$ correspondingly to
 ensure the resonance of the arm cavities.}
\label{tab:three_port}
\end{table}

\section{Case III: Achieving broadband sensitivity}
\label{sec:caseIII}

In the previous two sections, we have been focusing our investigation
on two specific intracavity filtering schemes; both offer explicit analytical
expressions that help to gain clear insights. However, these particular
models cover only a small parameter space of all possible intracavity
filtering schemes. In this section we present a numerical optimization with
the aim to maximize a certain cost function as to evaluate
the overall limits to the performance of this scheme.

\begin{figure}[!b]\centering
\includegraphics[width=0.48\textwidth]{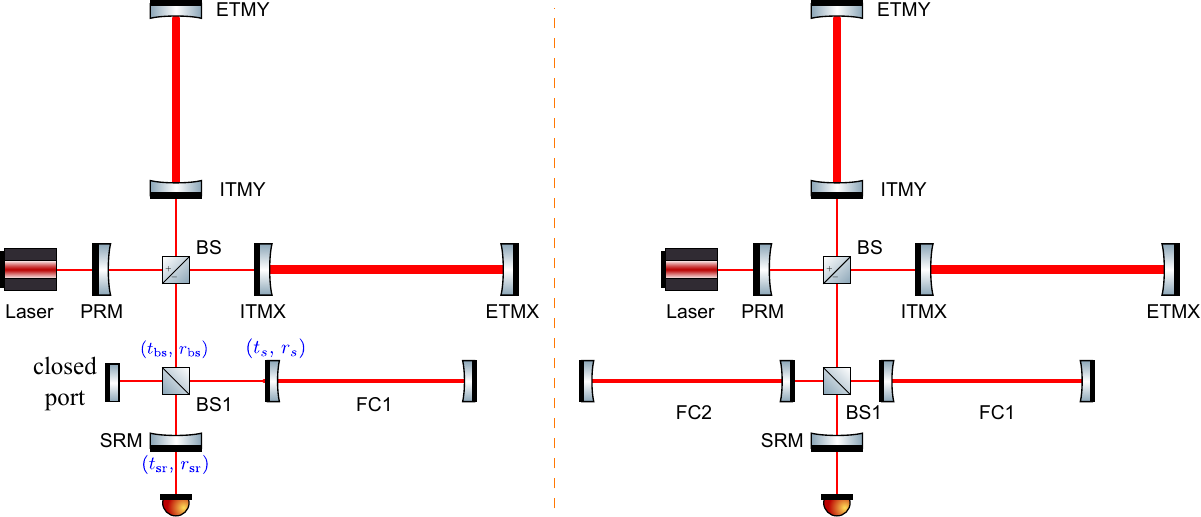}
\label{fig:intra_cavity_general_optimization_configuration}
\caption{The intracavity filtering schemes  with one filter cavity and a closed port (left) and two
filter cavities (right), for numerical optimization using the cost function
Eq.\,\eqref{eq:cost_function}.}
\label{fig:intra_cavity_general_optimization_configuration}
\end{figure}

For optimization, we use the cost function
introduced in~\cite{Miao13} including realistic nonquantum noises
(e.g., suspension and mirror coating thermal noise):
%and refer to Fig.\,\ref{fig:classical_noise} for the details):
\begin{equation}\label{eq:cost_function}
{\cal C}(\bm
x)=\left\{\int_{f_{\rm min}}^{f_{\rm max}} {\rm d}(\log_{10} f) \,
\log_{10} \left[\frac{h_{\rm ref}}{h_{\rm intra}(\bm
x)}\right]\right\}^{-1},
\end{equation}
where [$f_{\rm min}$, $f_{\rm max}$] is the frequency
range of the optimization; $\bm x$ is the set of optical parameters
that can be tuned by the algorithm, in particular
the parameters of the compound optics, including transmissivity and
reflectivity of the filter cavity, the SRM, and the BS1; $h_{\rm ref}$
is the square root of the total noise spectral density---the sum of 
quantum-noise and classical-noise spectral densities---of 
a reference design (which can
be the Advanced LIGO baseline design); and $h_{\rm intra}$ is the
square root of the total noise spectral density of the intracavity filtering scheme. We
will maximize the results by integrating over $\log
f$ instead of $f$ to give higher weight to low-frequency sensitivity.

\begin{figure}[!t]
\includegraphics[width=0.43\textwidth]{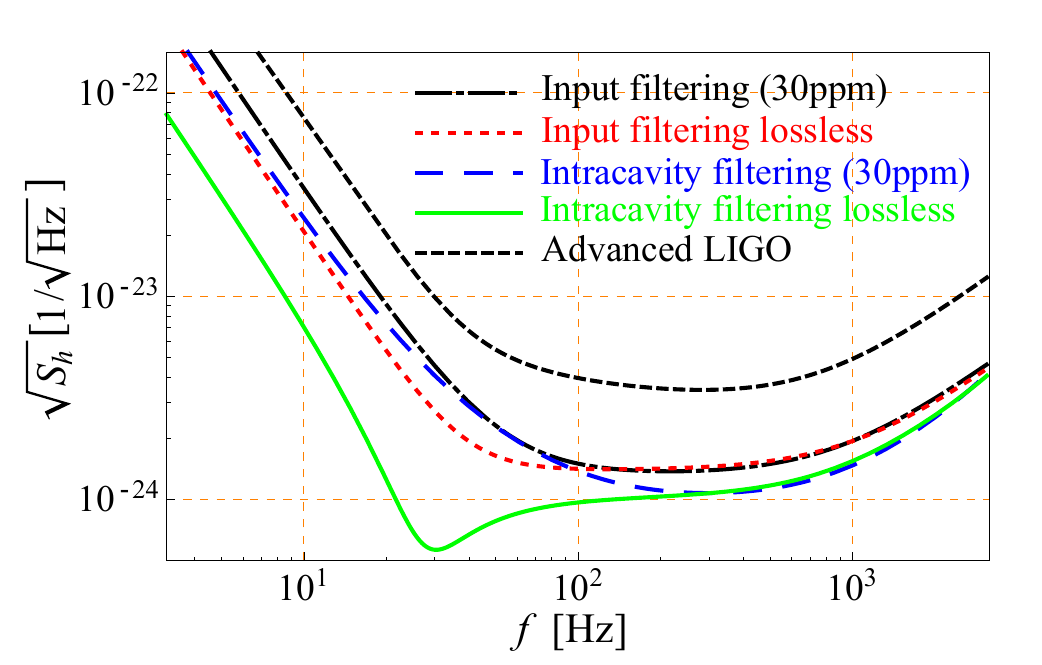}
\caption{Plot comparing the numerically optimized quantum-noise spectral
density of an intracavity filtering scheme [see
Fig.\,\ref{fig:intra_cavity_general_optimization_configuration}
(left) or Fig.\,\ref{fig:speed_meter_intra_cavity}] with an input-filtering scheme\,\cite{Kimble02}, 
which uses frequency-dependent squeezing to reduce quantum noise over a broad frequency band.
The quantum noise spectral density of Advanced LIGO is chosen as our reference\,\cite{Harry10}.}
\label{fig:intra_cavity_general_optimization_QN}
\end{figure}

\begin{figure}[!t]
\includegraphics[width=0.43\textwidth]{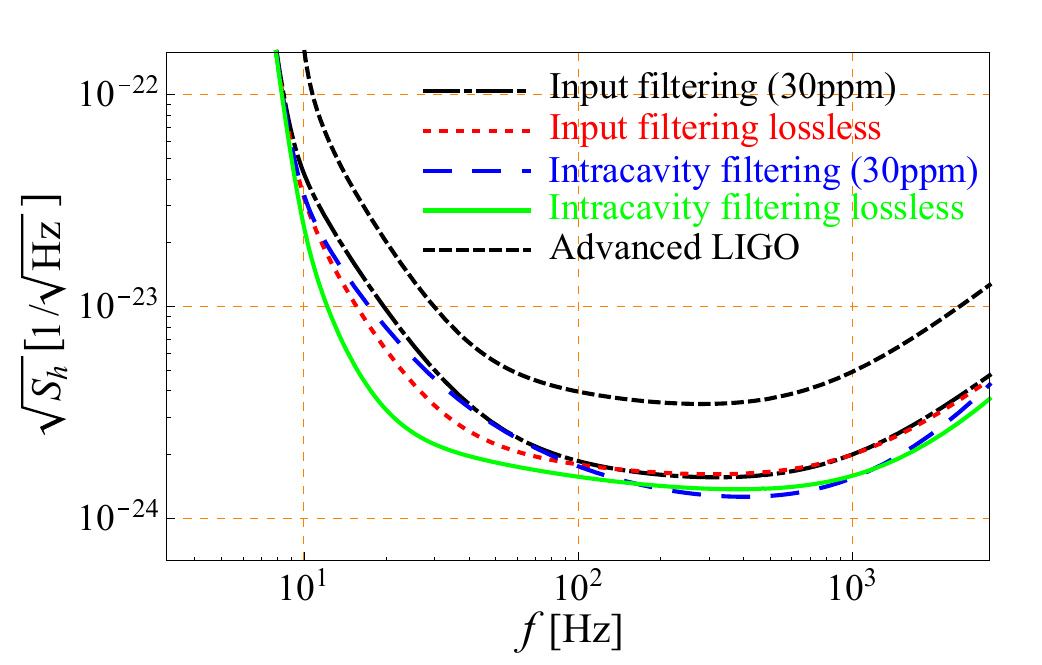}
\caption{Plot showing the numerically optimized total noise spectrum of an intracavity filtering scheme compared 
to an input-filtering case. Here classical noises such as thermal noise, gravity gradient noise and 
seismic noise are included based on the Advanced 
LIGO baseline design\,\cite{Harry10, LSC13}. }\label{fig:intra_cavity_general_optimization}
\end{figure}

\begin{table}[!b]
\begin{center}
\begin{tabular}{lcc}
\hline \hline
Parameter& Lossless case & Lossy case \\
\hline 
FC input mirror  & $0.0057\, (0.0)$ & $0.0090\,(0.0)$
\\ 
SRM & $0.040\,(-0.045)$ & $0.027\,(-0.018)$
\\ 
BS1 & $0.25\,(-0.017)$ & $0.13\,(-0.0032)$
\\
\hline
\end{tabular}
\end{center}
\caption{A table summarizing the optimal power transmissivity (and the reflection phase) of the optics in the
intracavity filtering scheme shown in
Fig.\,\ref{fig:intra_cavity_general_optimization_configuration}
(left). The lossless values refer to an idealized model. The 30\,ppm
column provides parameters based on a optical loss at the mirrors of
30\,ppm. The reflectivity
coefficients are complex numbers indicating the phase shift of the
propagation as shown in Fig.\,\ref{fig:three_port}.}
\label{tab:optima}
\end{table}

We initially considered an intracavity filtering scheme with two
optical cavities placed inside the signal-recycling cavity as shown in
the right-hand side of
Fig.\,\ref{fig:intra_cavity_general_optimization_configuration}. However, the optimization results suggested that a
configuration as shown in the left-hand side of
Fig.\,\ref{fig:intra_cavity_general_optimization_configuration} with
only one filter cavity is sufficient to achieve an equivalently good
result.  The outcomes of the optimization
are provided in
Fig.\,\ref{fig:intra_cavity_general_optimization_QN} and 
Fig.\,\ref{fig:intra_cavity_general_optimization}, where the dashed
black curve shows the noise spectral density of Advanced LIGO as a reference\,\cite{Harry10, LSC13}. The
optimized parameters are summarized in Table\,\ref{tab:optima}.
The intracavity filtering scheme with one optical cavity is able to reduce quantum
noise over a broadband as shown in Fig.\,\ref{fig:intra_cavity_general_optimization_QN}. 
For an ideal case, it is much better than an 
input-filtering scheme shown as the green solid curve against the red solid curve. 
In addition, we compare the
quantum noise of the ideal lossless case (green) with a model
including 30\,ppm mirror loss (blue). We found that optical losses
degrade low-frequency quantum noise in the intracavity filtering scheme and
its susceptibility to loss is similar to the input-filtering scheme which
is shown as the black solid curve in the figure 
(for both schemes, the filter cavity length is assumed to be 100m). Considering a 
30\,ppm mirror loss, the performance of an intracavity filtering scheme is slightly better than 
an input-filtering scheme. Figure\,\ref{fig:intra_cavity_general_optimization} compares the 
sensitivity of the input-filtering and intracavity filtering schemes respectively given that 
Advanced LIGO baseline design is applied with other classical 
noises, e.g., thermal noise, gravity gradient noise and seismic noise\,\cite{Harry10} being included.

\section{Conclusions}
\label{sec:conclusion} Previous work has shown that the quantum noise
of advanced gravitational-wave detectors
can be reduced over a broad frequency band by modifying the input
optics (input filtering) or output optics (output
filtering)~\cite{Kimble02}. We have investigated an alternative
filtering scheme---intracavity filtering, placing an optical cavity inside
the signal-recycling cavity, as a practical implementation for future
GW detectors.

We first considered the intracavity filtering as an alternative method to
cancel the radiation-pressure noise, hoping to reproduce the excellent low-frequency 
quantum-noise performance realized by the ideal frequency-dependent readout. 
However, it turned out that the filter cavity produces a frequency-dependent
phase shift that significantly reduced the sensitivity at intermediate frequencies.
More explicitly, we have shown the resulting noise spectrum:
(\rmnum{1}) at low frequencies, is scaled as $\sqrt{T_{\rm sr}}/\Omega^2$, (\rmnum{2})
at intermediate frequencies as $1/(\Omega\sqrt{T_{\rm sr}})$, and
(\rmnum{3}) at high frequencies as $\Omega/\sqrt{T_{\rm sr}}$.

We continued our investigation and considered the intracavity filtering as a
speed meter similar to the one proposed
in~\cite{Purdue02,Purdue02a}. Such a scheme, first
of all, does not require a RSE mirror. More important, it eases the stringent requirement of 
the sloshing-cavity design
presented in~\cite{Purdue02}, as the characteristic frequency for the speed
response is not determined by the sloshing mirror only, but is now also related to the
transmissivity of the arm cavity ITM ($T_{\rm ITM}$). In particular, the requirement on low transmissivity of the sloshing mirror can be relieved by a factor of $\sqrt{T_{\rm ITM}}$. We also found that the quantum noise of this
scheme, at low frequencies, shows a
position-meter-like response and then smoothly transits to a
speed-meter response in the intermediate frequencies.

The quantum-noise behavior of the intracavity filtering varies when choosing
different optical filters inside the signal-recycling cavity. We
numerically optimize the intracavity filtering scheme, aiming at reducing 
quantum noise over a broad frequency band. This optimization uses the 
Advanced LIGO sensitivity as a reference and assumed reduced classical 
noises (such as seismic noise, suspension and mirror thermal noise). We showed that, with reasonable optical losses,
for instance 30\,ppm per mirror, the quantum noise of an intracavity filtering scheme
is comparable to the frequency-dependent squeezing, so that this
scheme can be considered as an potential alternative.

In summary, even though an intracavity filter scheme is not able to completely evade
the radiation-pressure noise, we found its implementation as a speed
meter eases the tight requirements of the cavity design compared to a
sloshing-cavity speed meter. Meanwhile, its mixed position and speed
response encourages future investigation of an
intracavity filter scheme as a practical alternative for GW
detectors. Additionally, the global optimization of an intracavity
filtering produced a similarly low quantum-noise behavior as frequency-dependent
squeezing.

%%%%%% Acknowledgments %%%%%%

\section{Acknowledgements}

We would like to thank Farid Khalili, Stefan Danilishin, Sergey Vyatchanin and members in
the LIGO-MQM discussion group for discussions. We also thank Rana Adhikari, Matthew Evans,
Stefan Ballmer and members in the AIC working group for fruitful discussions. We
thank Frank Br\"{u}ckner for comments.
M.W. and A. F. have been supported by the Science and
Technology Facilities Council (STFC).  H. M. and Y. C.
are supported by the National Science Foundation (NSF) Grants
No. PHY-0555406, No. PHY-0956189, No. PHY-1068881, as well as the David and Barbara
Groce startup fund at Caltech, and the Institute for Quantum
Information and Matter, a Physics Frontier Center with funding from
the NSF and the Gordon and Betty Moore Foundation.
%%%%%% Appendix %%%%%%

\appendix

\section{Input-output relation for the three-port junction in the
intracavity filtering scheme}
\label{Appendix:A}

\begin{figure}[!b]
\includegraphics[width=0.48\textwidth]{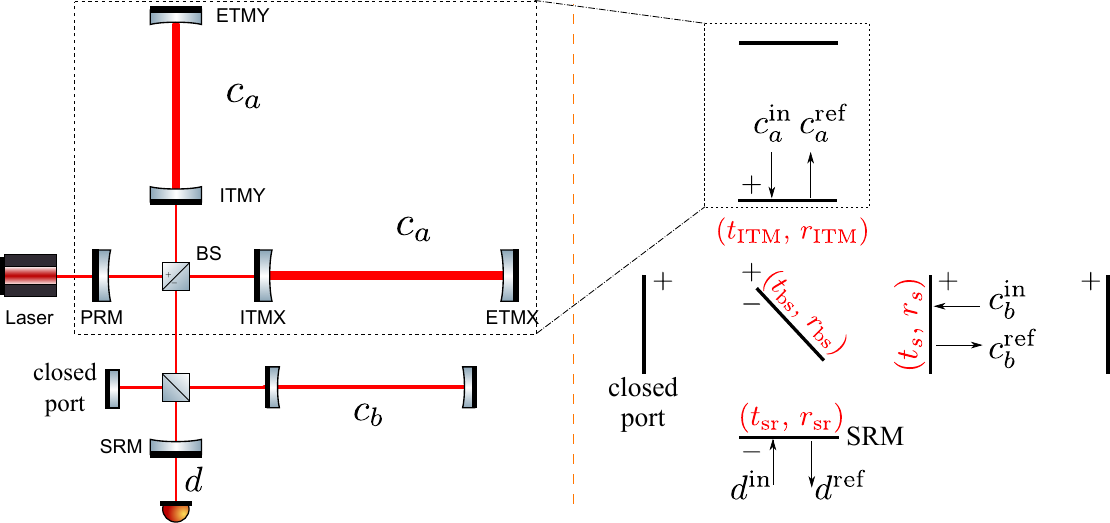}
\caption{Diagram illustrating the three-port junction that we are
interested in. We map the two arm cavities into a single cavity
(denoted by the dashed box), following~\cite{Buonanno03}. Here $r_i$
(complex) and $t_i$ (real) are the amplitude reflectivity and
transmissivity of the optics. The sign of convention for the
reflectivity are indicated by $\pm$---the amplitude reflectivity on
the positive side is $r$ and the minus side is $-r^*$ (complex
conjugate).
\label{fig:three_port}}
\end{figure}

In this section we provide details for the input-output relation
in the intracavity
filtering scheme and define the characteristic frequencies
used in the Hamiltonian in
Sec.\,\ref{sec:caseII}.

In Fig.\,\ref{fig:three_port}, we show the reflectivity for each
optics and its convention of sign. By using the junction condition on
each optics, we obtain
\begin{equation}\label{eq:input_output_three_port} \bm v^{\rm ref} =
{\bf M}_t\,\bm v^{\rm in}\,,
\end{equation} with $\bm v^{\rm ref}=[c_a^{\rm ref}, c_b^{\rm ref},
d^{\rm ref}]^{\rm T}$ and $\bm v^{\rm in}=[c_a^{\rm in}, c_b^{\rm in},
d^{\rm in}]^{\rm T}$ (the superscript $\rm T$ denotes the transpose)
and ${\bf M}_t$ being the transfer matrix. The transfer matrix has the
following property:
\begin{equation}\label{eq:stokes_relation} {\bf M}_t={\bf M}_t^{\rm
T},\quad {\bf M}_t\,{\bf M}_t^{\dag}={\bf I},
\end{equation} which means that ${\bf M}_t$ is a {\it symmetric
unitary} matrix. This gives the Stokes relation for such a three-port
linear optics. More specifically, the elements of ${\bf M}_t$ in terms
of the reflectivity and transmissivity of each optics are given by
\begin{align}\nonumber {\bf M}_{11}=&{\cal D}^{-1}[r_{\rm ITM}-r_s^*
r_{\rm bs}^2-r_{\rm ITM}r_{\rm sr}r_{\rm bs}^{*2}\\&\quad~~+(r_{\rm
sr}+r_{\rm ITM} r_s^*)t_{\rm bs}^2+r_s^* r_{\rm sr}]\,,\\ {\bf
M}_{12}=&{\cal D}^{-1}(r_{\rm bs}- r_{\rm sr} r_{\rm bs}^*)t_{\rm ITM}
\,t_s\,,\\ {\bf M}_{13}=&{\cal D}^{-1}(1+r_s^*)t_{\rm ITM}t_{\rm
sr}t_{\rm bs}\,,\\ \nonumber {\bf M}_{22}=&{\cal D}^{-1}[r_s-r_{\rm
ITM}^* r_{\rm bs}^2-r_sr_{\rm sr}r_{\rm bs}^{*2} \\&\quad~~+(1+r_{\rm
ITM}^*r_sr_{\rm sr})t_{\rm bs}^2+r_{\rm ITM}^* r_{\rm sr}]\,,\\ {\bf
M}_{23}=&{\cal D}^{-1}(-r_{\rm bs}^*-r_{\rm ITM}^* r_{\rm bs})t_s
t_{\rm sr}t_{\rm bs} \,,\\\nonumber {\bf M}_{33}=&{\cal
D}^{-1}[-r_{\rm sr}^*+r_{\rm bs}^{*2}+r_{\rm ITM}^*r_s^*r_{\rm
sr}^*r_{\rm bs}^2 \\&\quad~~-(r_{\rm ITM}^*+r_s^*r_{\rm sr}^*)t_{\rm
bs}^2-r_{\rm ITM}^*r_s^*]\,,
\end{align} where the denominator $\cal D$ reads
\begin{equation}\label{eq:denominator_D}\nonumber {\cal D}=1-r_{\rm
ITM}^*r_s^*r_{\rm bs}^2-r_{\rm sr}r_{\rm bs}^{*2}+(r_{\rm ITM}^*r_{\rm
sr}+r_s^*)t_{\rm bs}^2+r_{\rm ITM}^*r_s^*r_{\rm sr}\,.
\end{equation}

From the above input-output relation, we can define the effective
coupling among three optical modes $c_a, c_b$ and the external
continuum $d$. Specifically, we introduce the sloshing frequency
between $c_a$ and $c_b$:
\begin{equation}\label{eq:sloshing_frequency_definition}
\omega_{s}\equiv \frac{c|{\bf M}_{\rm 12}|}{2\sqrt{L_{\rm arm}L_s}}\,,
\end{equation} and two decay rates for each mode:
\begin{equation}\label{eq:decay_rate_definition} \gamma_a\equiv
\frac{c|{\bf M}_{\rm 13}|^2}{4L_{\rm arm}}\,,\;~\gamma_b\equiv
\frac{c|{\bf M}_{\rm 23}|^2}{4L_s}\,.
\end{equation} In addition, the resonant frequencies for each modes
can also be defined, and we have
\begin{equation}\label{eq:resonant_frequency-definition}
\omega_a\equiv \frac{c \,\arg({\bf M}_{11})}{2L_{\rm arm}}\,,\;~
\omega_b\equiv \frac{c\, \arg({\bf M}_{22})}{2L_1}\,,
\end{equation} with ``$\arg$'' being the phase angle.

%%%%%% References %%%%%%

\bibliography{intra}

\begin{thebibliography}{18}
\expandafter\ifx\csname natexlab\endcsname\relax\def\natexlab#1{#1}\fi
\expandafter\ifx\csname bibnamefont\endcsname\relax
  \def\bibnamefont#1{#1}\fi
\expandafter\ifx\csname bibfnamefont\endcsname\relax
  \def\bibfnamefont#1{#1}\fi
\expandafter\ifx\csname citenamefont\endcsname\relax
  \def\citenamefont#1{#1}\fi
\expandafter\ifx\csname url\endcsname\relax
  \def\url#1{\texttt{#1}}\fi
\expandafter\ifx\csname urlprefix\endcsname\relax\def\urlprefix{URL }\fi
\providecommand{\bibinfo}[2]{#2}
\providecommand{\eprint}[2][]{\url{#2}}

\bibitem[{\citenamefont{Harry and the LIGO
  Scientific~Collaboration}(2010)}]{Harry10}
\bibinfo{author}{\bibfnamefont{G.~M.} \bibnamefont{Harry}} \bibnamefont{and}
  \bibinfo{author}{\bibnamefont{the LIGO Scientific~Collaboration}},
  \bibinfo{journal}{Classical and Quantum Gravity}
  \textbf{\bibinfo{volume}{27}}, \bibinfo{pages}{084006}
  (\bibinfo{year}{2010}),
  \urlprefix\url{http://stacks.iop.org/0264-9381/27/i=8/a=084006}.

\bibitem[{\citenamefont{Accadia et~al.}(2011)\citenamefont{Accadia, Acernese,
  Antonucci, Astone, Ballardin, Barone, Barsuglia, Basti, Bauer, Bebronne
  et~al.}}]{Accadia11}
\bibinfo{author}{\bibfnamefont{T.}~\bibnamefont{Accadia}},
  \bibinfo{author}{\bibfnamefont{F.}~\bibnamefont{Acernese}},
  \bibinfo{author}{\bibfnamefont{F.}~\bibnamefont{Antonucci}},
  \bibinfo{author}{\bibfnamefont{P.}~\bibnamefont{Astone}},
  \bibinfo{author}{\bibfnamefont{G.}~\bibnamefont{Ballardin}},
  \bibinfo{author}{\bibfnamefont{F.}~\bibnamefont{Barone}},
  \bibinfo{author}{\bibfnamefont{M.}~\bibnamefont{Barsuglia}},
  \bibinfo{author}{\bibfnamefont{A.}~\bibnamefont{Basti}},
  \bibinfo{author}{\bibfnamefont{T.~S.} \bibnamefont{Bauer}},
  \bibinfo{author}{\bibfnamefont{M.}~\bibnamefont{Bebronne}},
  \bibnamefont{et~al.}, \bibinfo{journal}{Classical and Quantum Gravity}
  \textbf{\bibinfo{volume}{28}}, \bibinfo{pages}{114002}
  (\bibinfo{year}{2011}),
  \urlprefix\url{http://stacks.iop.org/0264-9381/28/i=11/a=114002}.

\bibitem[{\citenamefont{Chen}(2013)}]{Chen13}
\bibinfo{author}{\bibfnamefont{Y.}~\bibnamefont{Chen}},
  \bibinfo{journal}{Journal of Physics B: Atomic, Molecular and Optical
  Physics} \textbf{\bibinfo{volume}{46}}, \bibinfo{pages}{104001}
  (\bibinfo{year}{2013}),
  \urlprefix\url{http://stacks.iop.org/0953-4075/46/i=10/a=104001}.

\bibitem[{\citenamefont{Danilishin and Khalili}(2012)}]{Danilishin12}
\bibinfo{author}{\bibfnamefont{S.~L.} \bibnamefont{Danilishin}}
  \bibnamefont{and} \bibinfo{author}{\bibfnamefont{F.~Y.}
  \bibnamefont{Khalili}}, \bibinfo{journal}{Living Reviews in Relativity}
  \textbf{\bibinfo{volume}{15}} (\bibinfo{year}{2012}),
  \urlprefix\url{http://www.livingreviews.org/lrr-2012-5}.

\bibitem[{\citenamefont{Buonanno and Chen}(2001)}]{Buonanno01}
\bibinfo{author}{\bibfnamefont{A.}~\bibnamefont{Buonanno}} \bibnamefont{and}
  \bibinfo{author}{\bibfnamefont{Y.}~\bibnamefont{Chen}},
  \bibinfo{journal}{Phys. Rev. D} \textbf{\bibinfo{volume}{64}},
  \bibinfo{pages}{042006} (\bibinfo{year}{2001}),
  \urlprefix\url{http://link.aps.org/doi/10.1103/PhysRevD.64.042006}.

\bibitem[{\citenamefont{Purdue and Chen}(2002)}]{Purdue02}
\bibinfo{author}{\bibfnamefont{P.}~\bibnamefont{Purdue}} \bibnamefont{and}
  \bibinfo{author}{\bibfnamefont{Y.}~\bibnamefont{Chen}},
  \bibinfo{journal}{Phys. Rev. D} \textbf{\bibinfo{volume}{66}},
  \bibinfo{pages}{122004} (\bibinfo{year}{2002}),
  \urlprefix\url{http://link.aps.org/doi/10.1103/PhysRevD.66.122004}.

\bibitem[{\citenamefont{Wade et~al.}(2012)\citenamefont{Wade, McKenzie, Chen,
  Shaddock, Chow, and McClelland}}]{Wade12}
\bibinfo{author}{\bibfnamefont{A.~R.} \bibnamefont{Wade}},
  \bibinfo{author}{\bibfnamefont{K.}~\bibnamefont{McKenzie}},
  \bibinfo{author}{\bibfnamefont{Y.}~\bibnamefont{Chen}},
  \bibinfo{author}{\bibfnamefont{D.~A.} \bibnamefont{Shaddock}},
  \bibinfo{author}{\bibfnamefont{J.~H.} \bibnamefont{Chow}}, \bibnamefont{and}
  \bibinfo{author}{\bibfnamefont{D.~E.} \bibnamefont{McClelland}},
  \bibinfo{journal}{Phys. Rev. D} \textbf{\bibinfo{volume}{86}},
  \bibinfo{pages}{062001} (\bibinfo{year}{2012}),
  \urlprefix\url{http://link.aps.org/doi/10.1103/PhysRevD.86.062001}.

\bibitem[{\citenamefont{Gough and James}(2009)}]{Gough2007}
\bibinfo{author}{\bibfnamefont{J.}~\bibnamefont{Gough}} \bibnamefont{and}
  \bibinfo{author}{\bibfnamefont{M.}~\bibnamefont{James}},
  \bibinfo{journal}{Automatic Control, IEEE Transactions on}
  \textbf{\bibinfo{volume}{54}}, \bibinfo{pages}{2530} (\bibinfo{year}{2009}),
  ISSN \bibinfo{issn}{0018-9286},
  \urlprefix\url{http://ieeexplore.ieee.org/stamp/stamp.jsp?tp=&arnumber=5286277&isnumber=5306469}.

\bibitem[{\citenamefont{Mabuchi}(2008)}]{Mabuchi2008}
\bibinfo{author}{\bibfnamefont{H.}~\bibnamefont{Mabuchi}},
  \bibinfo{journal}{Phys. Rev. A} \textbf{\bibinfo{volume}{78}},
  \bibinfo{pages}{032323} (\bibinfo{year}{2008}),
  \urlprefix\url{http://link.aps.org/doi/10.1103/PhysRevA.78.032323}.

\bibitem[{\citenamefont{Kerckhoff et~al.}(2010)\citenamefont{Kerckhoff, Nurdin,
  Pavlichin, and Mabuchi}}]{Kerckhoff2010}
\bibinfo{author}{\bibfnamefont{J.}~\bibnamefont{Kerckhoff}},
  \bibinfo{author}{\bibfnamefont{H.~I.} \bibnamefont{Nurdin}},
  \bibinfo{author}{\bibfnamefont{D.~S.} \bibnamefont{Pavlichin}},
  \bibnamefont{and} \bibinfo{author}{\bibfnamefont{H.}~\bibnamefont{Mabuchi}},
  \bibinfo{journal}{Phys. Rev. Lett.} \textbf{\bibinfo{volume}{105}},
  \bibinfo{pages}{040502} (\bibinfo{year}{2010}),
  \urlprefix\url{http://link.aps.org/doi/10.1103/PhysRevLett.105.040502}.

\bibitem[{\citenamefont{Hamerly and Mabuchi}(2012)}]{Hamerly2012}
\bibinfo{author}{\bibfnamefont{R.}~\bibnamefont{Hamerly}} \bibnamefont{and}
  \bibinfo{author}{\bibfnamefont{H.}~\bibnamefont{Mabuchi}},
  \bibinfo{journal}{Phys. Rev. Lett.} \textbf{\bibinfo{volume}{109}},
  \bibinfo{pages}{173602} (\bibinfo{year}{2012}),
  \urlprefix\url{http://link.aps.org/doi/10.1103/PhysRevLett.109.173602}.

\bibitem[{\citenamefont{Kimble et~al.}(2001)\citenamefont{Kimble, Levin,
  Matsko, Thorne, and Vyatchanin}}]{Kimble02}
\bibinfo{author}{\bibfnamefont{H.~J.} \bibnamefont{Kimble}},
  \bibinfo{author}{\bibfnamefont{Y.}~\bibnamefont{Levin}},
  \bibinfo{author}{\bibfnamefont{A.~B.} \bibnamefont{Matsko}},
  \bibinfo{author}{\bibfnamefont{K.~S.} \bibnamefont{Thorne}},
  \bibnamefont{and} \bibinfo{author}{\bibfnamefont{S.~P.}
  \bibnamefont{Vyatchanin}}, \bibinfo{journal}{Phys. Rev. D}
  \textbf{\bibinfo{volume}{65}}, \bibinfo{pages}{022002}
  (\bibinfo{year}{2001}),
  \urlprefix\url{http://link.aps.org/doi/10.1103/PhysRevD.65.022002}.

\bibitem[{\citenamefont{Miao et~al.}(2013)\citenamefont{Miao, Yang, and
  andYanbei Chen}}]{Miao13}
\bibinfo{author}{\bibfnamefont{H.}~\bibnamefont{Miao}},
  \bibinfo{author}{\bibfnamefont{H.}~\bibnamefont{Yang}}, \bibnamefont{and}
  \bibinfo{author}{\bibfnamefont{R.~X.~A.} \bibnamefont{andYanbei Chen}},
  \bibinfo{journal}{arXiv: 1305.3957}  (\bibinfo{year}{2013}),
  \urlprefix\url{http://arxiv.org/abs/1305.3957}.

\bibitem[{\citenamefont{Braginsky and Khalilli}(1992)}]{Braginsky92}
\bibinfo{author}{\bibfnamefont{V.}~\bibnamefont{Braginsky}} \bibnamefont{and}
  \bibinfo{author}{\bibfnamefont{F.}~\bibnamefont{Khalilli}},
  \emph{\bibinfo{title}{Quantum Measurements}} (\bibinfo{publisher}{Cambridge
  University Press}, \bibinfo{year}{1992}),
  \urlprefix\url{http://dx.doi.org/10.1017/CBO9780511622748}.

\bibitem[{\citenamefont{Wang et~al.}(2013)\citenamefont{Wang, Bond, Brown,
  Br\"uckner, Carbone, Palmer, and Freise}}]{Wang13}
\bibinfo{author}{\bibfnamefont{M.}~\bibnamefont{Wang}},
  \bibinfo{author}{\bibfnamefont{C.}~\bibnamefont{Bond}},
  \bibinfo{author}{\bibfnamefont{D.}~\bibnamefont{Brown}},
  \bibinfo{author}{\bibfnamefont{F.}~\bibnamefont{Br\"uckner}},
  \bibinfo{author}{\bibfnamefont{L.}~\bibnamefont{Carbone}},
  \bibinfo{author}{\bibfnamefont{R.}~\bibnamefont{Palmer}}, \bibnamefont{and}
  \bibinfo{author}{\bibfnamefont{A.}~\bibnamefont{Freise}},
  \bibinfo{journal}{Phys. Rev. D} \textbf{\bibinfo{volume}{87}},
  \bibinfo{pages}{096008} (\bibinfo{year}{2013}),
  \urlprefix\url{http://link.aps.org/doi/10.1103/PhysRevD.87.096008}.

\bibitem[{\citenamefont{Buonanno and Chen}(2003)}]{Buonanno03}
\bibinfo{author}{\bibfnamefont{A.}~\bibnamefont{Buonanno}} \bibnamefont{and}
  \bibinfo{author}{\bibfnamefont{Y.}~\bibnamefont{Chen}},
  \bibinfo{journal}{Phys. Rev. D} \textbf{\bibinfo{volume}{67}},
  \bibinfo{pages}{062002} (\bibinfo{year}{2003}),
  \urlprefix\url{http://link.aps.org/doi/10.1103/PhysRevD.67.062002}.

\bibitem[{\citenamefont{{LIGO Scientific Collaboration}
  et~al.}(2013)\citenamefont{{LIGO Scientific Collaboration}, {Virgo
  Collaboration}, Aasi, Abadie, Abbott, Abbott, Abbott, Abernathy, Accadia,
  Acernese et~al.}}]{LSC13}
\bibinfo{author}{\bibnamefont{{LIGO Scientific Collaboration}}},
  \bibinfo{author}{\bibnamefont{{Virgo Collaboration}}},
  \bibinfo{author}{\bibfnamefont{J.}~\bibnamefont{Aasi}},
  \bibinfo{author}{\bibfnamefont{J.}~\bibnamefont{Abadie}},
  \bibinfo{author}{\bibfnamefont{B.~P.} \bibnamefont{Abbott}},
  \bibinfo{author}{\bibfnamefont{R.}~\bibnamefont{Abbott}},
  \bibinfo{author}{\bibfnamefont{T.~D.} \bibnamefont{Abbott}},
  \bibinfo{author}{\bibfnamefont{M.}~\bibnamefont{Abernathy}},
  \bibinfo{author}{\bibfnamefont{T.}~\bibnamefont{Accadia}},
  \bibinfo{author}{\bibfnamefont{F.}~\bibnamefont{Acernese}},
  \bibnamefont{et~al.}, p.~\bibinfo{pages}{18} (\bibinfo{year}{2013}),
  \eprint{1304.0670}, \urlprefix\url{http://arxiv.org/abs/1304.0670}.

\bibitem[{\citenamefont{Purdue}(2002)}]{Purdue02a}
\bibinfo{author}{\bibfnamefont{P.}~\bibnamefont{Purdue}},
  \bibinfo{journal}{Phys. Rev. D} \textbf{\bibinfo{volume}{66}},
  \bibinfo{pages}{022001} (\bibinfo{year}{2002}),
  \urlprefix\url{http://link.aps.org/doi/10.1103/PhysRevD.66.022001}.

\end{thebibliography}

%%%%%% End of the document %%%%%%

\end{document}